\documentclass[twocolumn,showpacs,prc,superscriptaddress]{revtex4-1}
\usepackage{graphicx,amsmath,amssymb,bm,multirow}
\usepackage{amsthm}
\usepackage{amsfonts}
\usepackage{dcolumn}
\usepackage{bm}
\usepackage{epstopdf}

\begin{document}
 \preprint{preprint}
\title{Shape evolution and coexistence in neutron-deficient Nd and Sm nuclei }
\author{J. Xiang}
\affiliation{School of Physics and Electronic, Qiannan Normal University for Nationalities, Duyun, 558000, China}

\author{Z. P. Li} \email{zpliphy@swu.edu.cn}
\affiliation{School of Physical Science and Technology, Southwest University, Chongqing 400715, China}
\affiliation{School of Physics and Electronic, Qiannan Normal University for Nationalities, Duyun, 558000, China}

\author{W. H. Long}
\affiliation{School of Nuclear Science and Technology, Lanzhou University, Lanzhou 730000, China}

\author{T. Nik\v si\' c}
\affiliation{Physics Department, Faculty of Science, University of Zagreb, Croatia}
\author{D. Vretenar}
\affiliation{Physics Department, Faculty of Science, University of Zagreb, Croatia}

\begin{abstract}
The evolution of shapes and low-energy shape coexistence is analyzed in neutron-deficient Nd and Sm nuclei, using a five-dimensional  quadrupole collective Hamiltonian (5DCH). Deformation energy surfaces, calculated with the relativistic energy density functional PC-PK1 and a separable finite-range pairing interaction, exhibit a transition from spherical shapes near $N=80$, to  $\gamma-$soft shapes, and to prolate deformed minima in lighter isotopes. The corresponding 5DCH model calculation, based on the self-consistent mean-field potentials,  reproduces the empirical isotopic trend of characteristic collective observables, and predicts  significantly different deformations for the first two $0^+$ states in the $N=74$ isotones $^{134}$Nd and $^{136}$Sm. In addition to  bands based on the triaxial $\gamma-$soft ground state, in excellent agreement with data, the occurrence of a  low-energy rotational band is predicted, built on the prolate deformed ($\beta\sim0.4$) excited state $0^+_2$.
\end{abstract}

\maketitle

\section{Introduction}
An interesting phenomenon that occurs in many mesoscopic systems is shape coexistence: sets of nearly degenerate low-energy  states are observed that can be characterized by different geometrical shapes. In atomic nuclei, in particular,  coexistence of different shapes in a single nucleus, and shape transitions as a function of nucleon number, have been observed in light, medium-heavy, and heavy systems~\cite{Heyde1983PR,Heyde2011RMP,Wood1992PR}. The distinctive character of shape coexistence in  nuclei reflects the interplay between single-particle and collective degrees of freedom. Experimental and theoretical studies of shapes, their evolution and transitions, provide  crucial information on the origin of nuclear collectivity and modification of shell structures in nuclei far from stability~\cite{Wrzosek-Lipska2016JPG}. In a recent systematic analysis of characteristic signatures of coexisting nuclear shapes in different mass regions \cite{Quan2017PRC}, quadrupole shape invariants for  more than six hundred even-even nuclei were calculated using a global self-consistent theoretical method based on universal energy density functionals (EDFs) and the quadrupole collective model. A  systematic comparison of shape invariants for the two lowest $0^+$ states has identified regions of possible shape coexistence. Different geometric shapes at low energies emerge as a universal structure property that occurs in different mass regions over the entire chart of nuclides.

Neutron-deficient rare-earth nuclei in the mass $A\approx140$ region are known to exhibit a variety of coexisting structures resulting from the mid-shell filling of the $h_{11/2}$ intruder orbital. Protons fill the lower part of the h$_{11/2}$ shell with low-$\Omega$ orbitals favoring a prolate nuclear deformation, whereas for neutrons the Fermi surface lies close to high-$\Omega$ h$_{11/2}$ states, favoring oblate shapes. This competition between opposing trends for proton and neutron deformations can produce $\gamma$-soft ground-state shapes with a significant degree of triaxiality. Calculations by M\"{o}ller and Bengtsson\cite{MOller2008ADNDT}, based on the finite-range droplet model, predicted  the ground states of neutron-deficient $A\sim140$ nuclei to be triaxial. In addition, several studies indicated that the $\gamma$ degree of freedom plays an important role in the description of collective excited states\cite{Carlsson2008PRC,Liu2008CPL,Mertz2008PRC,Starzecki1988PLB,Muller-Veggian1980NPA}. These nuclei exhibit additional interesting structure effects, with the observation of isomeric states\cite{Starzecki1988PLB,Muller-Veggian1978NPA,Yoshikawa1975NPA}, and magnetic rotation\cite{Lieder2002EPJA,Sugawara2009PRC}. Based on experimental results, it has been  suggested that $^{144}$Dy\cite{Procter2010PRC} and $^{142}$Sm\cite{Rajbanshi2014PRC} exhibit possible shape coexistence. Furthermore, in Ref.\cite{Quan2017PRC}, medium-deformed triaxial ground states coexisting with highly deformed prolate excited state have been predicted in this region, especially for the nuclei $^{134}$Nd, $^{136,138}$Sm, $^{140,142}$Gd, and $^{142,144}$Dy.

Rare-earth nuclei with neutron number $N \approx 90$ present some of the best examples of  shape phase transitions. Employing a consistent framework of structure models based on  energy density functionals, in several studies we analysed microscopic signatures of ground-state shape phase transitions in this region of the nuclear mass table. In the present work we consider  shape evolution and the possible occurrence of shape coexistence in neutron-deficient Nd and Sm isotopes.  The analysis starts from self-consistent mean-field calculations of deformation energy surfaces using relativistic energy density functionals \cite{Ring1996PPNP,Vretenar2005PR,Meng2006PPNP,Meng2016}, and is extended to include the treatment of collective correlations with the five-dimensional collective Hamiltonian (5DCH)  model. This approach has successfully been applied to the description of low-lying collective states in a wide range of nuclei,  from the mass region $A\sim40$ to superheavy systems \cite{Niksic2009PRC,Li2009PRCa,Li2009PRCb,Li2010PRC,Li2011PRC,Li2012PLB,Fu2013PRC,Lu2015PRC,Li2016JPG,Niksic2011PPNP,Prassa2012PRC,Prassa2013PRC,Xiang2016PRC}.

In Sec.~\ref{method} we present a short outline of the theoretical framework used to study shape coexistence in neutron-deficient $A\sim140$ nuclei.  The systematics of collective deformation energy surfaces, the evolution of characteristic signatures of deformed shapes, and the low-energy spectra of $^{134}$Nd and $^{136}$Sm are discussed in Sec.~\ref{Results}. Section \ref{Summary} summarizes the principal results.

\section{\label{method} The 5D collective Hamiltonian}

Nuclear EDF-based studies of low-energy structure phenomena start from a  self-consistent mean-field (SCMF) calculation of  deformation energy surfaces with mass multipole moments as constrained quantities.  The results are static symmetry-breaking product many-body states.  The static nuclear mean-field is characterised by the  breaking of symmetries of the underlying Hamiltonian -- translational, rotational, particle number and, therefore, includes static correlations, e.g. deformations and  pairing. To calculate excitation spectra and electromagnetic transition rates it is necessary to extend the SCMF scheme to include collective correlations that arise from symmetry restoration and fluctuations around the mean-field minima.

Low-energy excitation spectra and transitions can be described using a collective Hamiltonian, with deformation-dependent parameters determined from microscopic SCMF calculations. For instance, in the case of quadrupole degrees of freedom,  excitations determined by quadrupole vibrational and rotational degrees of freedom can be described by considering two quadrupole collective coordinates $\beta,\gamma$ and three Euler angles $\Omega\equiv(\phi,\theta,\psi)$ \cite{Niksic2009PRC}. The corresponding 5DCH Hamiltonian takes the following form,
\begin{equation}\label{5DCH}
 \hat{H} (\beta, \gamma,\Omega) =\hat{T}_{\text{vib}}+\hat{T}_{\text{rot}} +V_{\text{coll}},
\end{equation}
where $V_{\text{coll}}$ is the collective potential that includes zero-point energy (ZPE) corrections, and  $\hat{T}_{\text{vib}}$  and $\hat{T}_{\text{rot}}$ are the vibrational and rotational kinetic energy terms, respectively,  \cite{Libert1999PRC,Prochniak2004NPA,Niksic2009PRC}

 \begin{align}
\hat{T}_{\text{vib}}=&-\frac{\hbar^2}{2\sqrt{wr}} \left\{\frac{1}{\beta^4}  \left[\frac{\partial}{\partial\beta}\sqrt{\frac{r}{w}}\beta^4  B_{\gamma\gamma} \frac{\partial}{\partial\beta}\right.\right.\nonumber\\
& \left.\left.- \frac{\partial}{\partial\beta}\sqrt{\frac{r}{w}}\beta^3 B_{\beta\gamma}\frac{\partial}{\partial\gamma} \right]+\frac{1}{\beta\sin{3\gamma}} \left[    -\frac{\partial}{\partial\gamma} \right.\right.\\
& \left.\left.\sqrt{\frac{r}{w}}\sin{3\gamma}  B_{\beta \gamma}\frac{\partial}{\partial\beta} +\frac{1}{\beta}\frac{\partial}{\partial\gamma} \sqrt{\frac{r}{w}}\sin{3\gamma} B_{\beta \beta}\frac{\partial}{\partial\gamma} \right]\right\}\nonumber,\\
\hat{T}_{\text{\text{\text{rot}}}} =&\frac{1}{2}\sum_{k=1}^3{\frac{\hat{J}^2_k}{\mathcal{I}_k}}.
\end{align}
$\hat{J}_k$ denotes the components of the angular momentum in the body-fixed frame of a nucleus, and $B_{\beta\beta}$, $B_{\beta\gamma}$, $B_{\gamma\gamma}$,  are the mass parameters. Two additional quantities that appear in $\hat T_{\text{vib}}$, namely $r=B_1B_2B_3$ (see Ref.~\cite{Niksic2009PRC} for the definition of $B_k$), and $w=B_{\beta\beta}B_{\gamma\gamma}-B_{\beta\gamma}^2 $, determine the volume element in the collective space. The constrained SCMF solutions for the  single-quasiparticle energies and wave functions for the entire energy surface, as functions of the quadrupole deformations $\beta$ and $\gamma$, provide the microscopic input for  calculation of the mass parameters, moments of inertia and the collective  potential. The Hamiltonian describes quadrupole vibrations, rotations, and the coupling of these collective modes. The dynamics of the 5DCH is governed by the seven functions of the intrinsic deformations $\beta$ and $\gamma$: the collective potential $V_{\rm coll}(\beta, \gamma)$, three mass parameters $B_{\beta\beta}(\beta, \gamma)$, $B_{\beta\gamma}(\beta, \gamma)$\ and $B_{\gamma\gamma}(\beta, \gamma)$, and three moments of inertia $\mathcal{I}_k(\beta, \gamma)$ ($k=1, 2, 3$). The corresponding eigenvalue equation  is solved by expanding the eigenfunctions on a complete set of basis functions that depend on the deformation variables $\beta$ and $\gamma$, and the Euler angles \cite{Prochniak2004NPA}.

\section{\label{Results}Neutron-deficient Nd and Sm isotopes}

Just as our recent global analysis of quadrupole shape invariants \cite{Quan2017PRC}, the present study starts from a quadrupole deformation-constrained relativistic mean-field (RMF) plus BCS calculation, with the point-coupling energy functional PC-PK1~\cite{Zhao2010PRC} and a separable pairing interaction~\cite{Tian2009PLB} in the particle-hole and particle-particle channels, respectively. The SCMF single-nucleon Dirac equation is solved by expanding the Dirac spinor in terms of a 3D harmonic oscillator basis with $14$ major shells. The SCMF states are calculated on the grid: $\beta\in[0.0, 0.8]$ and $\gamma\in[0^\circ, 60^\circ]$, with $\Delta\beta=0.05$ and $\Delta\gamma=6^\circ$. More details about the mean-field calculations can be found in Refs.~\cite{Niksic2010PRC} and \cite{Xiang2012NPA}. For the collective Hamiltonian (\ref{5DCH}) the mass parameters and moments of inertia are determined in the perturbative cranking approximation using the SCMF triaxial quasi-particle states~\cite{Niksic2009PRC}. Diagonalization of the 5DCH generates the  excitation spectra and collective wave functions that are used to calculate  spectroscopic properties, such as electric multipole transition strengths \cite{Niksic2009PRC}.

\begin{figure}[t]
\includegraphics[width=0.45\textwidth]{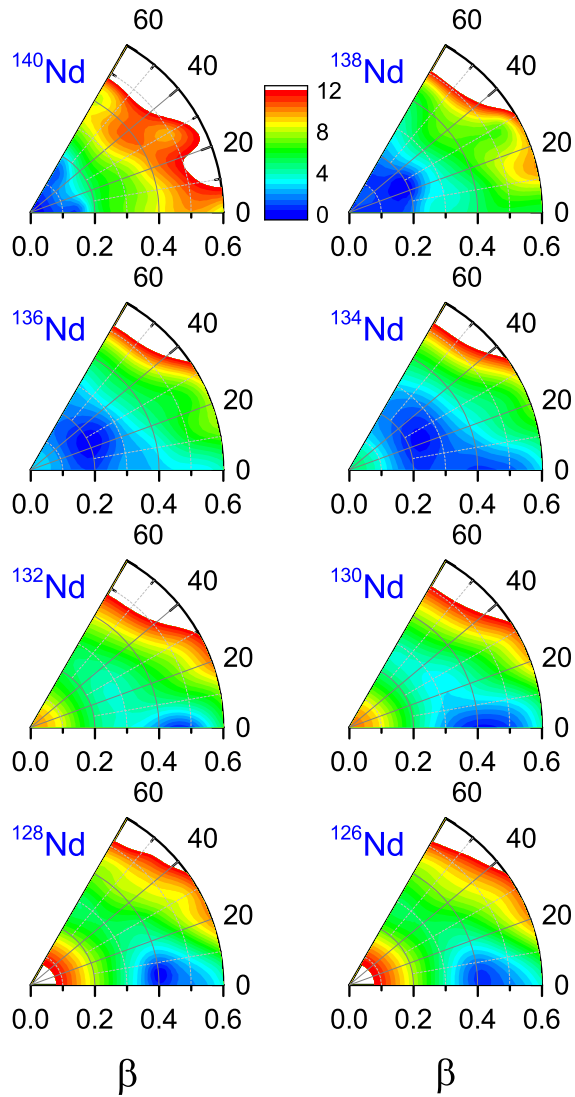}
\caption{\label{Vcol-Nd}(Color online) Collective potential energy V$_{\text{coll}}$ of the even-even isotopes $^{126-140}$Nd in the $(\beta, \gamma)$ plane,
obtained by constrained relativistic mean-field (RMF) plus BCS calculations. For each nucleus energies are normalized to the absolute minimum. The energy difference between neighboring contours is $0.5$ MeV.}
\end{figure}

\begin{figure}[t]
\includegraphics[width=0.45\textwidth]{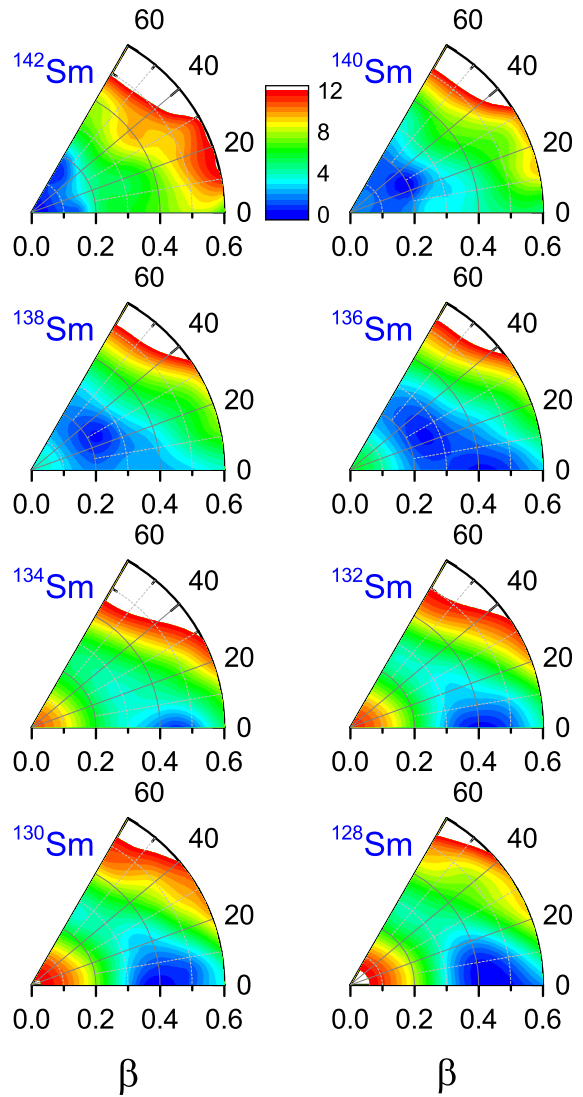}
\caption{\label{Vcol-Sm}(Color online) Same as in the caption to Fig.~\ref{Vcol-Nd} but for the neutron-deficient isotopes of Sm.}
\end{figure}

Fig. \ref{Vcol-Nd} displays the collective potential energy surface in the $\beta-\gamma$ plane for even-even neutron-deficient isotopes $^{126-140}$Nd. For each nucleus the calculated energies are normalized with respect to the binding energy of the absolute minimum. The corresponding deformation energy maps of the even-even isotopes  $^{128-142}$Sm are shown in Fig. \ref{Vcol-Sm}. For both isotopic chains these plots illustrate a rapid transition from spherical shapes near $N=80$, to $\gamma-$soft shapes, and then to prolate deformed energy surfaces in lighter isotopes. Starting from the spherical nuclei $^{140}$Nd and $^{142}$Sm, a certain degree of triaxiality develops in $^{138}$Nd and $^{140}$Sm, followed by the occurrence of $\gamma-$soft minima in $^{136}$Nd and $^{138}$Sm. At neutron number $N=74$, both Nd and Sm exhibit a coexistence of $\gamma-$soft and axially-deformed prolate shapes. For $N\leq72$ isotopes only well-deformed axially-symmetric prolate minima are predicted by the SCMF calculation.

\begin{figure}[h]
\includegraphics[width=0.45\textwidth]{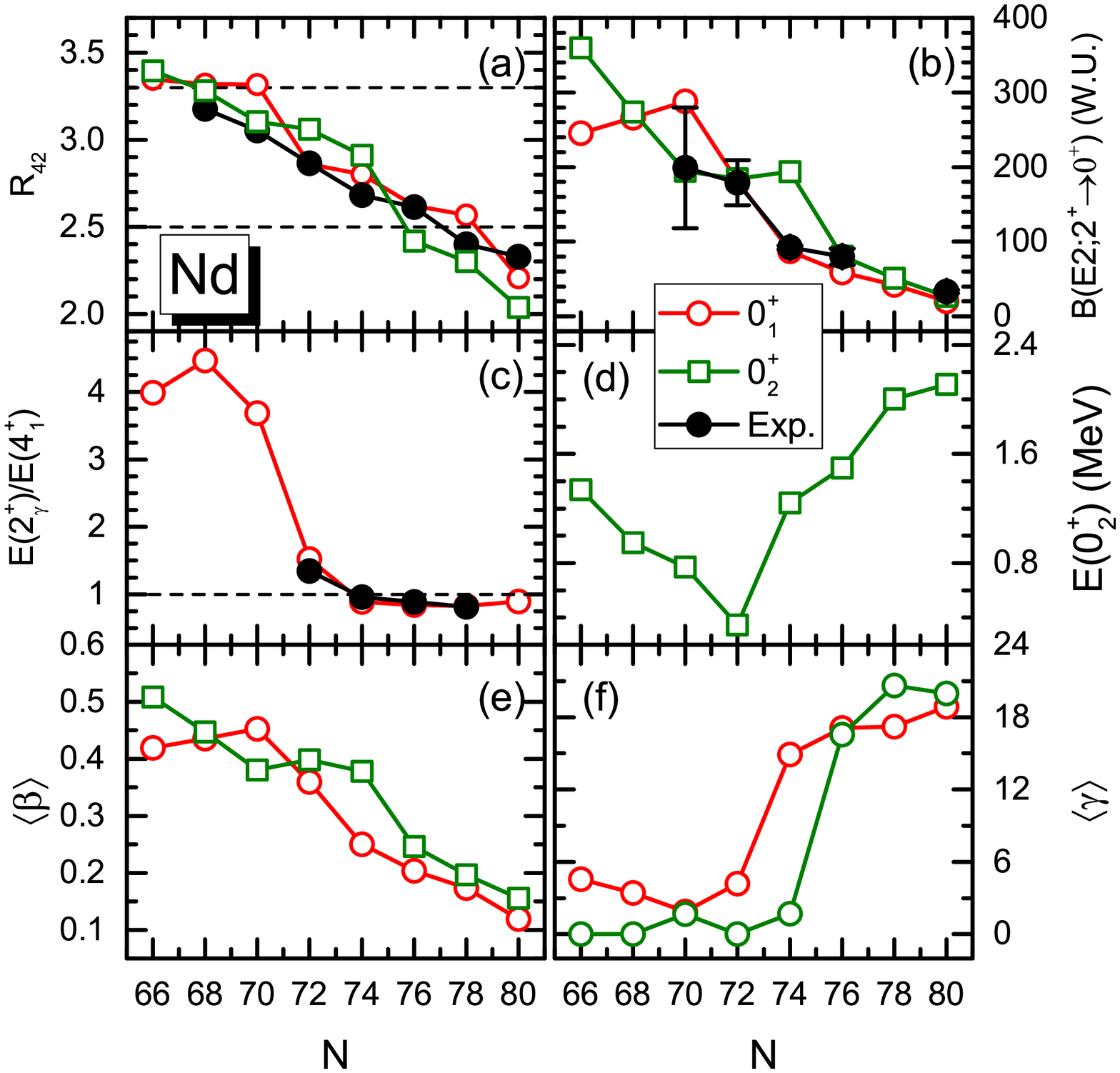}
\caption{\label{Obs-Nd}(Color online)  The energy ratio $R_{42}=\frac{E(4^+)-E(0^+)}{E(2^+)-E(0^+)}$ (a) and $B(E2;2^+\rightarrow0^+)$ values (b) for the ground-state band and the band based on the $0^+_2$ state, the energy ratio $E(2^+_\gamma)/E(4^+_1)$ (c), the excitation energy of the $0^+_2$ state $E(0^+_2)$ (d), the values of the quadrupole deformation parameters $\beta$ (e) and $\gamma$ (f) for the $0^+_1$ and $0^+_2$ state, as functions of the neutron number in Nd isotopes. The data are from Ref.~\cite{NNDC}.}
\end{figure}

\begin{figure}[h]
\includegraphics[width=0.45\textwidth]{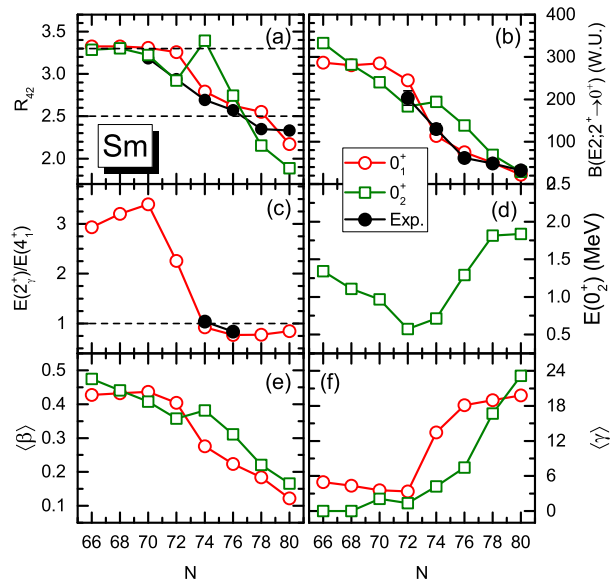}
\caption{\label{Obs-Sm}(Color online) Same as in the caption to Fig. \ref{Obs-Nd} but for the chain of Sm isotopes.}
\end{figure}

In Figs. \ref{Obs-Nd} and \ref{Obs-Sm} we analyze the evolution of several quantities that can be used to characterize transitions between different shapes as functions of the  neutron number: the energy ratio $R_{42}=\frac{E(4^+)-E(0^+)}{E(2^+)-E(0^+)}$ and the $B(E2;2^+\rightarrow0^+)$ values for ground-state band and the band based on $0^+_2$, the energy ratio $E(2^+_\gamma)/E(4^+_1)$, the excitation energy of the $0^+_2$ state, and the quadrupole deformation parameters $\beta$ and $\gamma$ in the ground state $0^+_1$ and the first excited $0^+_2$. For the even-even Nd and Sm isotopes these quantities are calculated using the 5DCH,  with parameters determined by the SCMF solutions shown in Figs.~\ref{Vcol-Nd} and \ref{Vcol-Sm}. Where available, the results are shown in comparison to data\cite{NNDC}. The evolution of $R_{42}$ characterizes shape transitions between axially-deformed rotors ($R_{42}=3.33$), $\gamma$-soft shapes ($R_{42}=2.50$),  and spherical vibrational nuclei ($R_{42}=2.00$). One notices how the energy ratio $R_{42}$ for the ground-state band decreases from the rotational value $R_{42}\sim3.3$ at  $N \sim 66$ to the transitional $\gamma-$soft value $R_{42}\sim2.5$ in the region $N=74-78$, in excellent agreement with the empirical trend. A similar behavior, with the  significant exception of $^{136}$Sm, is also predicted for the sequence of levels built on the first excited $0^+$ state.  As shown in Figs.\ref{Obs-Nd}(c) and \ref{Obs-Sm}(c), the ratio $E(2^+_\gamma)/E(4^+_1)$ calculated with the 5DCH in Nd and Sm isotopes is in excellent agreement with the available data. The predicted values $\sim 1.0$ for this energy ratio at $N=74-78$ characterize the occurrence of low-energy $\gamma$-deformed structures in these isotopes.

Figures \ref{Obs-Nd}(b) and \ref{Obs-Sm}(b) display the values $B(E2;2^+\rightarrow0^+)$ for the transitions from the first state $2^+_1$ to the ground-state, and for the state  $2^+$ built on the excited $0^+_2$ state, in the Nd and Sm isotopes, respectively. The corresponding experimental values  $B(E2;2^+_1\rightarrow0^+_1)$ \cite{NNDC} are shown for comparison.  The 5DCH calculation reproduce the data, and one notices the gap between the value $B(E2)$ for the ground-state sequence and the corresponding value for the state  built on the $0^+_2$ state, in $^{134}$Nd and $^{136,138}$Sm. This indicates that in these isotopes the average  deformations $\langle\beta\rangle$ for the ground state $0^+_1$ differ from  those of the $0^+_2$ state, as also shown in Fig. \ref{Obs-Nd}(e) and Fig. \ref{Obs-Sm}(e). The energies of the first excited $0^+$ state of Nd and Sm isotopes are shown in Figs. \ref{Obs-Nd}(d) and \ref{Obs-Sm}(d), respectively. For both chains $E(0^+_2)$ decreases from $N=66$ to $N=72$, where it displays a pronounced minimum, and then increases sharply toward the neutron closed shell at $N=82$. This behavior clearly reflects the evolution of the collective potential energy surfaces from prolate deformed to $\gamma$-soft, and to spherical shapes, as shown in Figs.~\ref{Vcol-Nd} and \ref{Vcol-Sm}.

The occurrence of shape coexistence can be further illustrated by analyzing the evolution of average quadrupole deformations  $\langle\beta\rangle$ and $\langle\gamma\rangle$ for the  two lowest $0^+$ states. In Figs.~\ref{Obs-Nd} and \ref{Obs-Sm} we plot the deformations $\langle\beta\rangle$ and $\langle\gamma\rangle$  for $0^+_1$ and $0^+_2$, as functions of neutron number in Nd and Sm isotopes, respectively. The average deformations are determined from the calculated quadrupole shape invariants following the procedure described in  Ref.~\cite{Quan2017PRC}. Note that, while they exhibit similar trends, marked differences between  $0^+_1$ and $0^+_2$ are predicted in several nuclei, especially at $N=74$. In $^{134}$Nd the values of $\langle\beta\rangle$ for the $0^+_1$ and $0^+_2$ states  are $\sim0.25$ and $\sim0.4$, and the corresponding $\langle\gamma\rangle$ are $\sim15^\circ$ and $2^\circ$, respectively. Therefore, not only the SCMF potential  energy surface, but also the 5DCH model calculation predicts a coexistence of soft triaxial and prolate axially-deformed low-energy structures in $^{134}$Nd. A very similar  picture is also found in $^{136}$Sm.

\begin{figure}[]
 \includegraphics[width=0.5\textwidth]{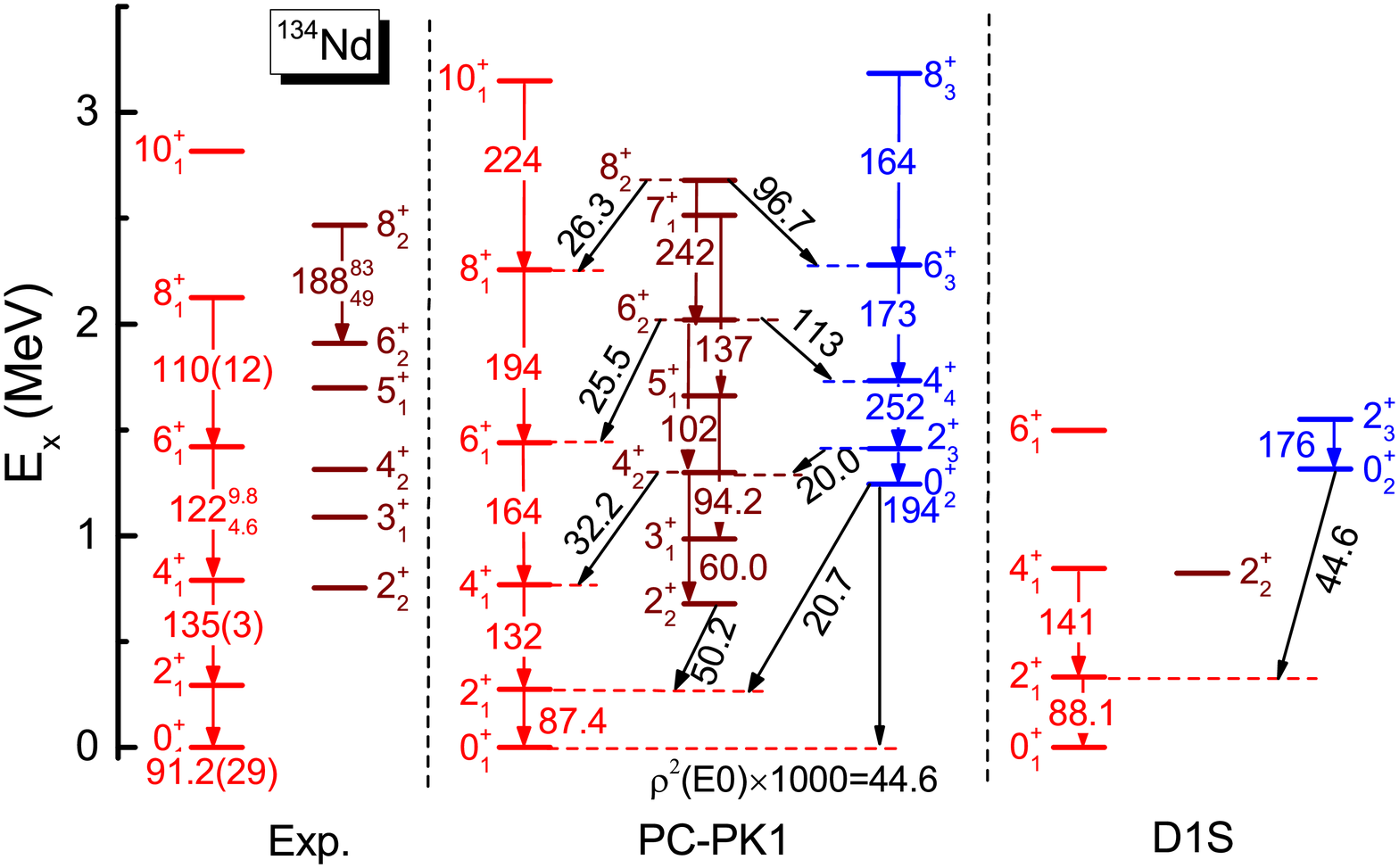}
 \includegraphics[width=0.5\textwidth]{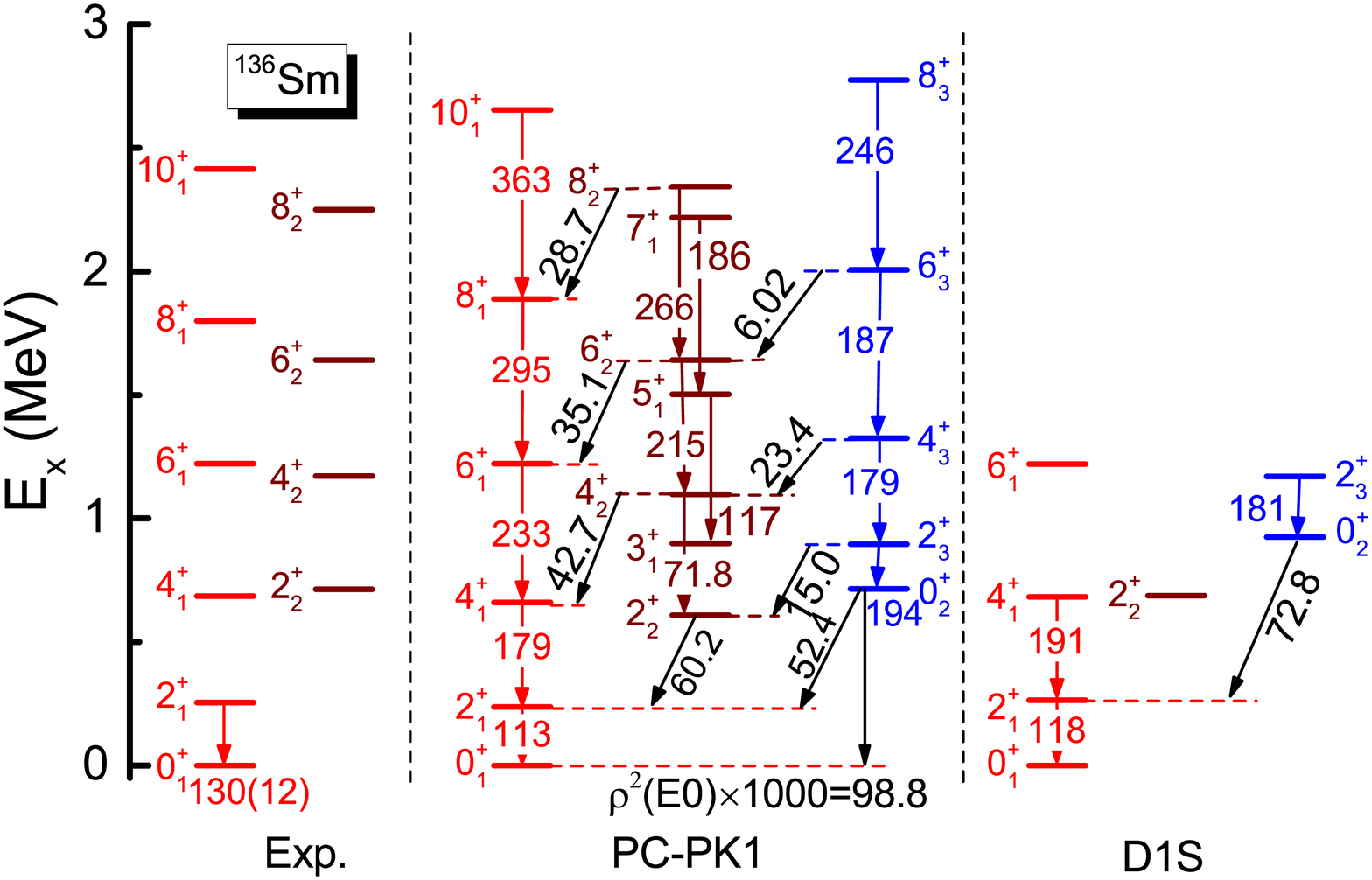}
\caption{\label{spec-CeNd}(Color online) 5DCH energy spectra of $^{134}$Nd and $^{136}$Sm (PC-PK1), compared to available data ~\cite{NNDC,Klemme1999PRC} (Exp.),
and to results obtained with the 5DCH based on the Gogny force (D1S) \cite{Delaroche2010}.}
\end{figure}

For these two nuclei in Fig.~\ref{spec-CeNd} we display the 5DCH low-energy spectra obtained in the present study (PC-PK1) and with the Gogny force  (D1S) \cite{Delaroche2010}, in comparison with available data \cite{NNDC,Klemme1999PRC}. Both models, without any additional adjustment,  reproduce the excitation energies and E2-transition rates of the two lowest bands based on the $\gamma-$soft ground state. In particular, we note that the  very low band-head $2^+_\gamma$ of the gamma-band is calculated slightly below $4^+_1$ in $^{134}$Nd, in excellent agreement with experiment.  Both 5DCH calculations predict the occurrence of a rotational sequence of levels in $^{134}$Nd and $^{136}$Sm, based on the low-lying $0^+_2$ state, that is,  structures built on the prolate deformed minima shown in Figs.~\ref{Vcol-Nd} and \ref{Vcol-Sm}. Compared to the yrast sequences, these bands are characterized by  larger moments of inertia and enhanced E2 transition rates for the few lowest levels. For higher angular momenta one expects considerable mixing between the two structures. The calculated E0 transition strengths $\rho^2(E0; 0^+_2\to 0^+_1)\times 10^3$ are: $44.6$ for $^{134}$Nd, and $98.8$ for $^{136}$Sm, comparable to the values that characterize the well-known shape-coexisting nuclei $^{98}$Sr [51(5)] and $^{100}$Zr  [108(19)] \cite{Kibedi2005ADNDT}. The bands built on the $0^+_2$ states have not yet been observed and it would, therefore, be very important to be able to  experimentally confirm the predicted shape coexistence in these $N=74$ isotones.

\begin{figure}[h]
\includegraphics[width=0.45\textwidth]{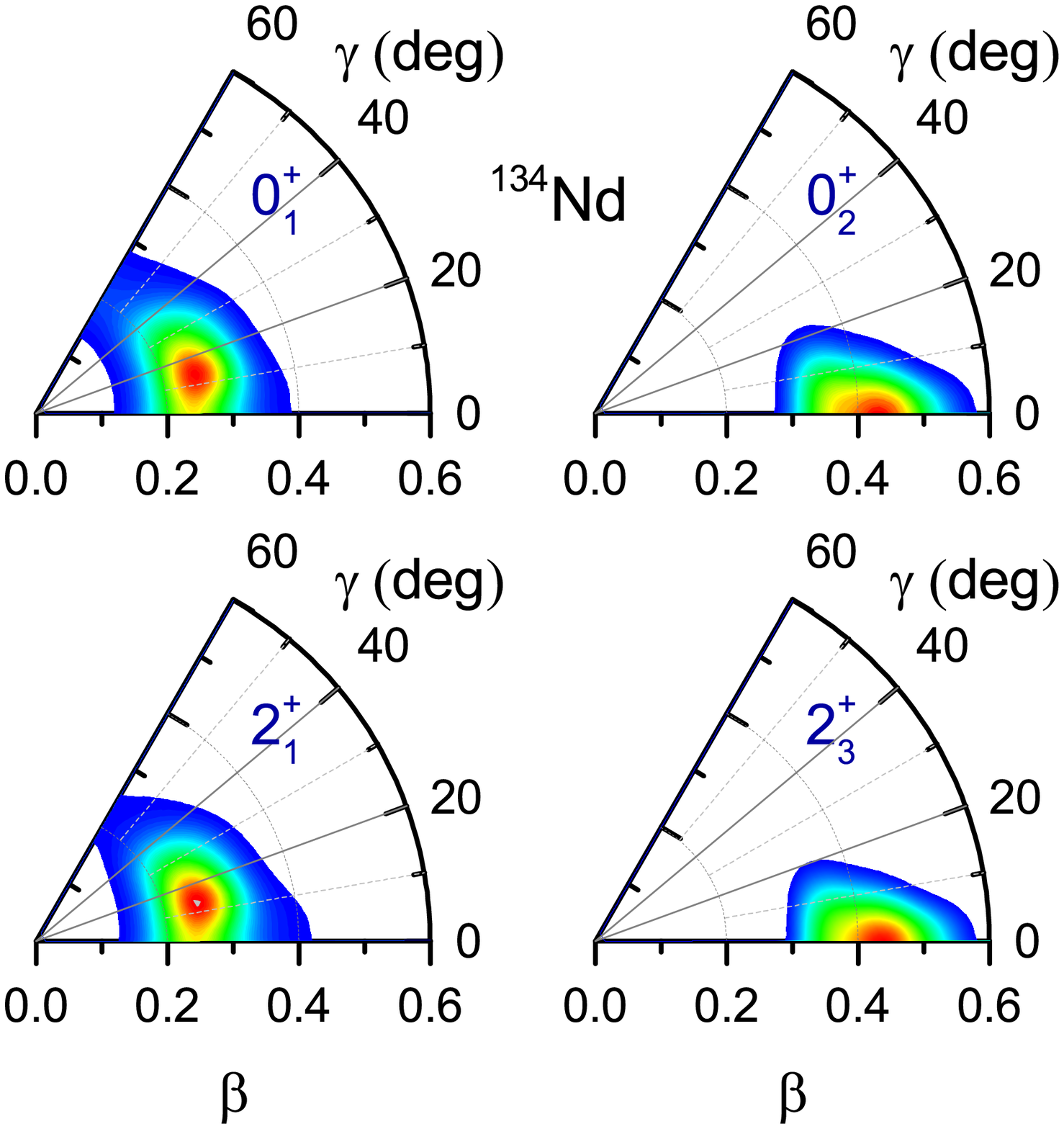}
\includegraphics[width=0.45\textwidth]{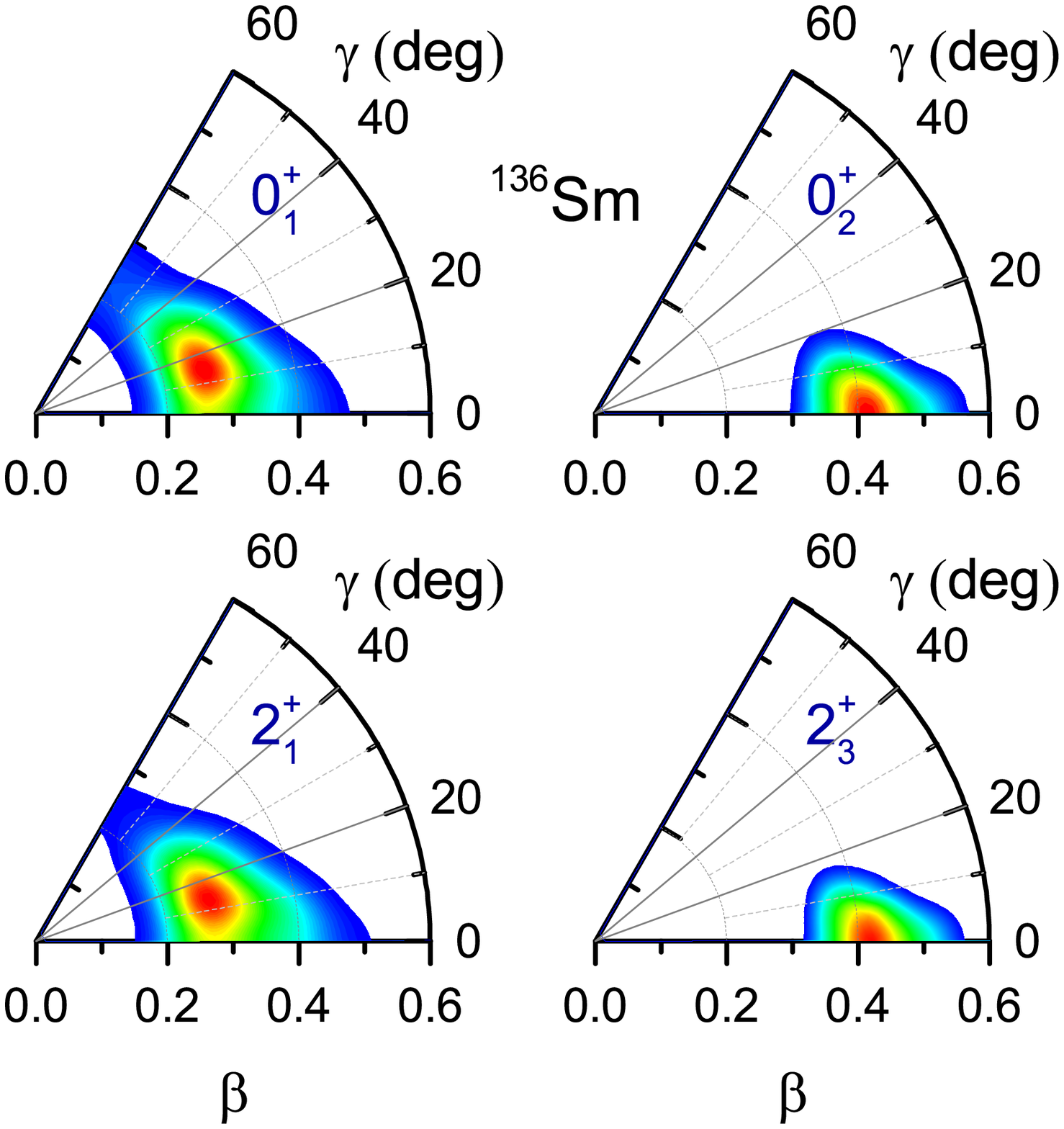}
\caption{\label{wav-NdSm}(Color online) Distribution of the probability density $\rho_{J\alpha}(\beta,\gamma)$ Eq.~(\ref{eq:probability}) for $0^+_1$, $0^+_2$, $2^+_1$ and $2^+_3$
collective states of $^{134}$Nd and $^{136}$Sm.}
\end{figure}

Shape coexistence in the 5DCH model is best illustrated by considering the probability density distributions that correspond to the collective wave functions in the $\beta-\gamma$ plane. The eigenfunctions of the collective Hamiltonian read
\begin{equation}
\label{wave-coll}
\Psi_\alpha^{JM}(\beta,\gamma,\Omega) =
  \sum_{K\in \Delta J}
           {\psi_{\alpha K}^J(\beta,\gamma)\Phi_{MK}^J(\Omega)}\; .
\end{equation}
For a given collective state, the probability distribution in the $(\beta,\gamma)$ plane is defined as
\begin{equation}
\rho_{J\alpha}(\beta,\gamma) = \sum_{K \in \Delta J}{ \left| \psi_{\alpha K}^J(\beta,\gamma)\right|^2\beta^3 },
\label{eq:probability}
\end{equation}
with the summation over the allowed set of values of the projection $K$ of the angular  momentum $J$ on the body-fixed symmetry axis, and  with the normalization
\begin{equation}
\int_0^\infty{\beta d\beta \int_0^{2\pi}{ \rho_{J\alpha}(\beta,\gamma)|\sin{3\gamma}|d\gamma }}=1.
\end{equation}
Figure~\ref{wav-NdSm} displays the distribution of probability density $\rho_{J\alpha}(\beta,\gamma)$ in the $\beta-\gamma$ plane for the two lowest  levels of the yrast sequence and the prolate band based on $0^+_2$, for $^{134}$Nd and $^{136}$Sm. These distributions clearly show that the two lowest collective states  correspond to a triaxial but $\gamma-$soft geometric shape, whereas the collective wave functions of the states $0^+_2$ and $2^+_3$ are concentrated on the prolate axis at considerably larger axial deformation $\beta\sim0.4$.

\begin{figure}[h]
\includegraphics[width=0.5\textwidth]{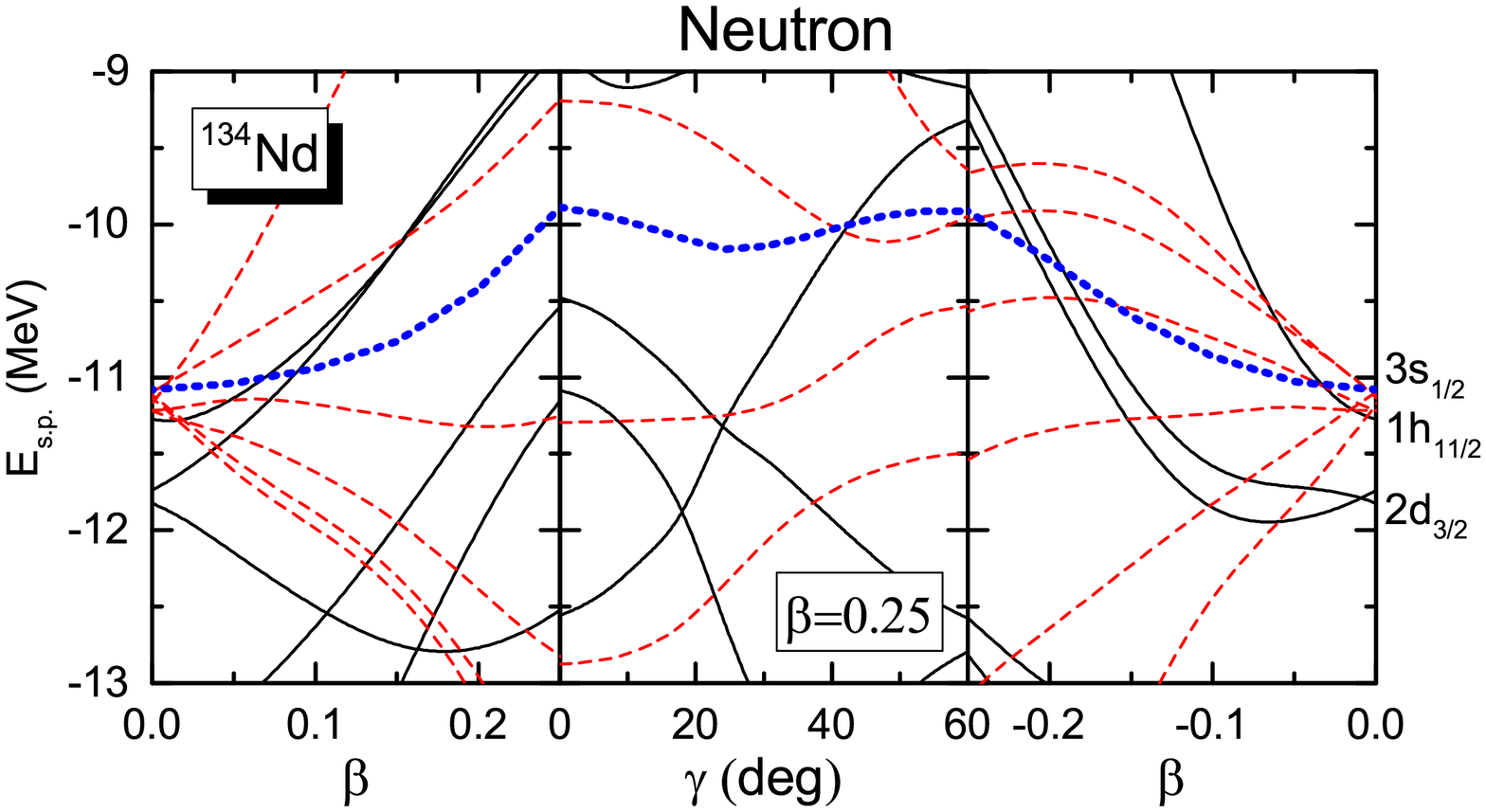}\\
\includegraphics[width=0.5\textwidth]{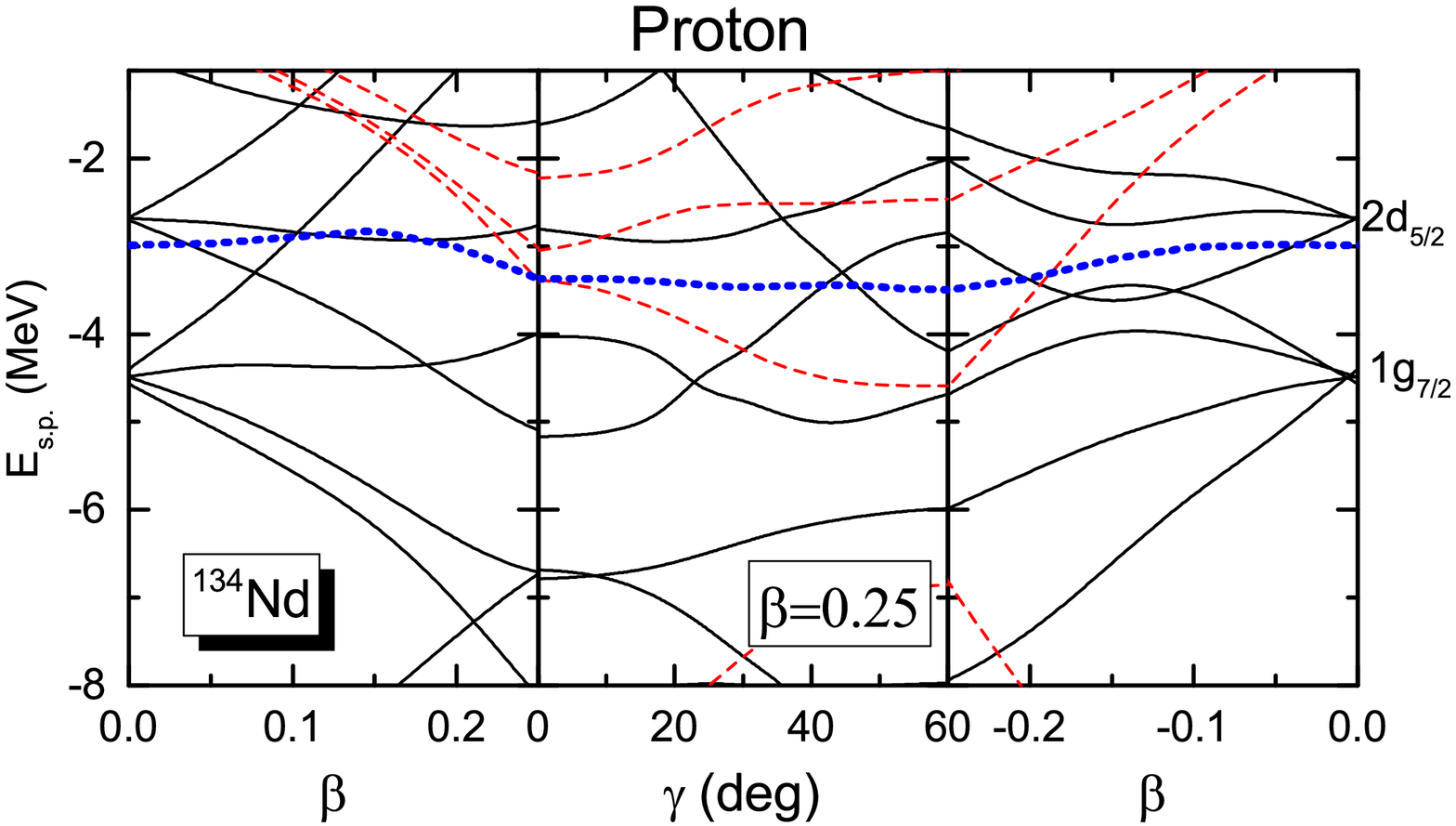}
\caption{\label{SPE-gam}(Color online) Neutron (upper panel) and proton (lower panel)
single-nucleon energy levels of $^{134}$Nd, as functions of the
deformation parameters along a closed path in the $\beta - \gamma$ plane.
Solid (black) curves correspond to levels with positive parity, and
 dashed (red) curves denote negative-parity levels.
The dotted (blue) curves correspond to the Fermi levels.
The panels on the left and right display prolate ($\gamma =0^\circ$) and oblate
 ($\gamma =60^\circ$) axially-symmetric single-particle levels, respectively.
 In the middle panel the proton and neutron levels are
 plotted as functions of $\gamma$ for a fixed value
 $|\beta|=0.25$, corresponding to the approximate position of the mean-field minimum.}
\end{figure}

\begin{figure}[h]
\includegraphics[width=0.5\textwidth]{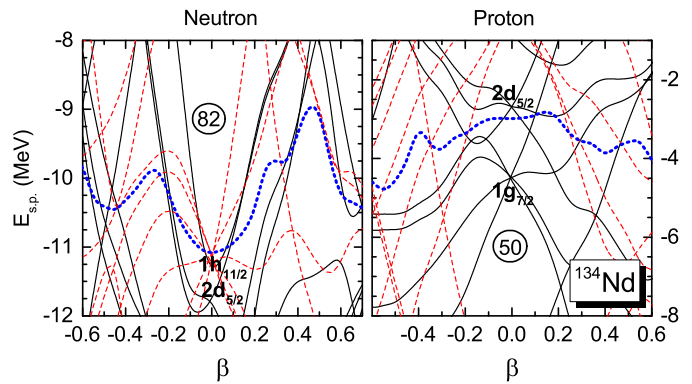}
\caption{\label{SPE-bet}(Color online) Single-neutron and single-proton levels in $^{134}$Nd as functions of the axial deformation parameter $\beta$. Solid (black) and dashed (red) curves denote the positive- and negative-parity levels, respectively. The dotted (blue) curves are the corresponding Fermi levels.}
\end{figure}

\begin{figure}[h]
\includegraphics[width=0.5\textwidth]{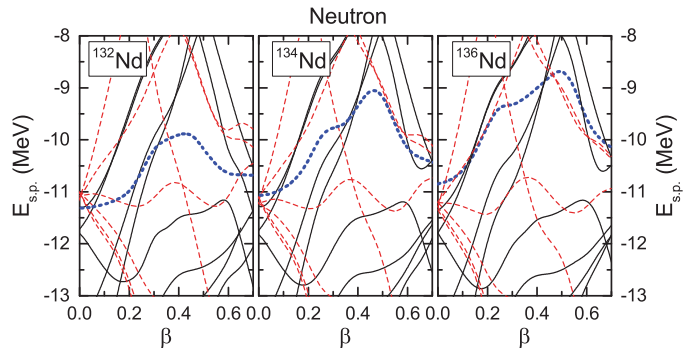}
\caption{\label{SPE-N}(Color online) Single-neutron levels in $^{132, 134, 136}$Nd isotopes as functions of the axial deformation parameter $\beta$. Solid (black) and dashed (red) curves correspond to the positive- and negative-parity levels, respectively. The dotted (blue) curves denote the Fermi levels.}
\end{figure}

The rapid transition from triaxial-soft to prolate shape around $N \approx 74$ can be understood from a microscopic point of view by considering the evolution of neutron and proton single-particle levels as functions of deformation and particle number. Namely, the formation of deformed minima is related to the occurrence of regions of low single-particle level density around the Fermi surface. In Fig.~\ref{SPE-gam} we plot the neutron (upper panel) and proton (lower panel) single-particle levels of $^{134}$Nd, as functions of the deformation parameters along a closed path in the $\beta - \gamma$ plane. Solid (black) curves correspond to levels with positive parity, and dashed (red) curves denote negative-parity levels. The dotted (blue) curves correspond to the Fermi levels.
Starting from the spherical configuration, we follow the single-nucleon levels on a  path along the prolate axis up to the approximate position of the mean-field minimum (left panel), then for this fixed value of $\beta$ the path from $\gamma = 0^\circ$ to $\gamma=60^\circ$ (middle panel) and, finally, back to the spherical configuration along the oblate axis (right panel). Configurations along the oblate axis are denoted by negative values of $\beta$. One notices that both neutron and proton levels display pronounced gaps between the last occupied and first unoccupied states in the triaxial region close to $\gamma\sim20^\circ$ (neutrons) and $\gamma\sim30^\circ$ (protons). These gaps give rise to the triaxial $\gamma$-soft  minimum shown in Fig. \ref{Vcol-Nd} (cf. also the distribution of the probability density for the states $0^+_1$ and $2^+_1$ in Fig.~\ref{wav-NdSm}). The origin of the low-lying prolate minimum in $^{134}$Nd, on which the band based on $0^+_2$ is built (cf. the states $0^+_2$ and $2^+_3$ in Figure~\ref{wav-NdSm}), can be found in the large energy gap for the proton levels at  $\beta \approx 0.4$, as shown in Fig.~\ref{SPE-bet}. By decreasing the number of neutrons by two, one reaches $^{132}$Nd. The proton levels do not change much, of course, whereas the neutron Fermi level  is lowered in energy. As a result the pronounced triaxial gap present in the single-neutron spectrum of $^{134}$Nd (cf. Fig.~\ref{SPE-gam}) disappears, while a large gap is formed among the axial single-neutron levels at $\beta \approx 0.5$. This is illustrated in Fig.~\ref{SPE-N}, where we plot the single-neutron levels in $^{132, 134, 136}$Nd as functions of the axial deformation parameter $\beta$. Together with the proton gap at $\beta \approx 0.4$, the large gap exhibited by the neutron levels leads to the formation of the prolate equilibrium minimum in $^{132}$Nd, as shown in Fig. \ref{Vcol-Nd}.

\section{\label{Summary} Summary}
We have analyzed the evolution of shapes and possible occurrence of low-energy shape coexistence in neutron-deficient Nd and Sm nuclei. Fully self-consistent mean-field triaxial calculations, based on the relativistic energy density functional PC-PK1 and a separable finite-range pairing interaction, have been performed to produce deformation energy surfaces in the $(\beta, \gamma)$ plane. These surfaces display a transition from spherical shapes near $N=80$, to $\gamma-$soft shapes, and eventually to pronounced prolate deformed minima in lighter isotopes. In particular, the Nd and Sm $N=74$ isotones exhibit coexisting low-energy $\gamma-$soft and axially-deformed prolate minima.

The SCMF deformation-constrained solutions provide a microscopic input for the parameters of the 5D quadrupole collective Hamiltonian that has been used to calculate spectroscopic properties of low-energy states. The 5DCH model calculation reproduces the empirical isotopic trend of the characteristic collective observables $R_{42}=\frac{E(4^+)-E(0^+)}{E(2^+)-E(0^+)}$, $B(E2;2^+\rightarrow0^+)$, and $E(2^+_\gamma)/E(4^+_1)$, while the values of $\langle\beta\rangle$, $\langle\gamma\rangle$ for the first two $0^+$ states indicate significantly different deformations of these states in $^{134}$Nd and $^{136}$Sm. The theoretical low-energy collective spectra of these two nuclei, including excitation energies and E2 transition rates, are in excellent agreement with the available data. In addition to the bands based on the triaxial $\gamma-$soft ground state, the model predicts the occurrence of a low-energy rotational band built on the prolate deformed ($\beta\sim0.4$) excited state $0^+_2$. $^{134}$Nd and $^{136}$Sm therefore present a very nice example of coexisting triaxial and prolate deformed shapes at low energy in neutron deficient rare-earth nuclei.

\begin{acknowledgements}
This work has been supported in part by the NSFC under Grants No. 11765015, No. 11475140, No. 11875225, No. 11675065, Joint Fund Project of Education Department in Guizhou Province(No. Qian Jiao He KY Zi[2016]312), Qiannan normal University Initial Research Foundation Grant to Doctor(qnsyrc201617), the Foundation of Scientific Innovative Research Team of Education Department of Guizhou Province (201329), and by the QuantiXLie Centre of Excellence, a project co-financed by the Croatian Government and European Union through the European Regional Development Fund - the Competitiveness and Cohesion Operational Programme (KK.01.1.1.01).
\end{acknowledgements}


\begin{thebibliography}{99}
\bibitem{Heyde2011RMP}K. Heyde and J. L. Wood, Rev. Mod. Phys. 83, 1467 (2011).
\bibitem{Heyde1983PR}K. Heyde, P. V. Isacker, M. Waroquier, J. L. Wood, and R. A. Meyer, Phys. Rep. 102, 291 (1983).
\bibitem{Wood1992PR}J. L. Wood, K. Heyde, W. Nazarewicz, M. Huyse, and P. V. Duppen, Phys. Rep. 215, 101 (1992).
\bibitem{Wrzosek-Lipska2016JPG}K. Wrzosek-Lipska and L. P. Gaffney, J. Phys. G: Nucl. Part. Phys. 43, 024012(2016).
\bibitem{Quan2017PRC}S. Quan, Q. Chen, Z. P. Li, T. Nik\v{s}i\'{c} and D. Vretenar, Phys. Rev. C 95, 054321 (2017).
\bibitem{MOller2008ADNDT}P. M\"{o}ller, R. Bengtsson, B. G. Carlsson, P. Olivius, T. Ichikawa, H. Sagawa, and A. Iwamoto, At. Data Nucl. Data Tables 94, 758(2008).
\bibitem{Starzecki1988PLB}W. Starzecki, G. DeAngelis, B. Rubio, J. Styczen, K. Zuber, H. G\"{u}ven, W. Urban, W. Gast, P. Kleinheinz, S. Lunardi, F. Soramel, A. Facco, C. Signorini, M. Morando, W. Meczynski, A. M. Stefanini, and G.Fortuna, Phys. Lett. B 200, 419(1988).
\bibitem{Carlsson2008PRC}B. G. Carlsson, I. Ragnarsson, R. Bengtsson, E. O. Lieder, R. M.
Lieder, and A. A. Pasternak, Phys. Rev. C 78, 034316 (2008).
\bibitem{Muller-Veggian1980NPA}M. M\"{u}ller-Veggian, H. Beuscher, D. R. Haenni, R. M. Lieder, A. Neskakis, and C.Mayer-B\"{o}ricke, Nucl. Phys. A 344, 89 (1980).

\bibitem{Liu2008CPL}H. L. Liu and  F. R. Xu, Chin. Phys. Lett. 25, 1621(2008).
\bibitem{Mertz2008PRC}A. F. Mertz, E. A. McCutchan, R. F. Casten, R. J. Casperson, A. Heinz, B. Huber, R. L\"{u}ttke, J. Qian, B. Shoraka, J. R. Terry, V. Werner, E. Williams, and R. Winkler, Phys. Rev. C 77, 014307(2008).
\bibitem{Muller-Veggian1978NPA}M. M\"{u}ller-Veggian, Y.Gono, R. M. Lieder, A. Neskakis, and C. Mayer-B\"{o}ricke, Nucl. Phys. A 304, 1(1978).
\bibitem{Yoshikawa1975NPA}N. Yoshikawa, Nucl. Phys. A 243, 143(1975).
\bibitem{Lieder2002EPJA}R. M. Lieder, T. Rzaca-Urban, H. Brands, W. Gast, H. M. J\"{a}ger, L. Mihailescu, Z. Marcinkowska, W. Urban, T. Morek, Ch. Droste, P. Szyma\'{n}ski, S. Chmel, D. Bazzacco, G. Falconi, R. Menegazzo, S. Lunardi, C. Rossi  Alvarez, G. de Angelis, E. Farnea, A. Gadea, D.R. Napoli, Z. Podolyak, Ts. Venkova, and R. Wyss, Eur. Phys. J. A 13, 297(2002).
\bibitem{Sugawara2009PRC}M. Sugawara, Y. Toh, M. Oshima, M. Koizumi, A. Osa, A. Kimura, Y. Hatsukawa, J. Goto, H. Kusakari, T. Morikawa, Y. H. Zhang, X. H. Zhou, Y. X. Guo, and M. L. Liu, Phys. Rev. C 79, 064321(2009).
\bibitem{Procter2010PRC}M. G. Procter, D. M. Cullen, C. Scholey, B. Niclasen, P. J. R. Mason, S. V. Rigby, J. A. Dare, A. Dewald, P. T. Greenlees, H. Iwasaki, U. Jakobsson, P. M. Jones, R. Julin, S. Juutinen, S. Ketelhut, M. Leino, N. M. Lumley,
O. M\"{o}ller, M. Nyman, P. Peura, T. Pissulla, A. Puurunen, P. Rahkila, W. Rother, P. Ruotsalainen, J. Sar\'{e}n, J. Sorri, and J. Uusitalo, Phys. Rev. C 81, 054320 (2010).
\bibitem{Rajbanshi2014PRC}S. Rajbanshi, A. Bisoi, S. Nag, S. Saha, J. Sethi, T. Trivedi, T. Bhattacharjee, S. Bhattacharyya, S. Chattopadhyay, G. Gangopadhyay, G. Mukherjee, R. Palit, R. Raut, M. Saha Sarkar, A. K. Singh, and A. Goswami, Phys. Rev. C 89, 014315(2014).

\bibitem{Ring1996PPNP} P. Ring, Prog. Part. Nucl. Phys. 37, 193 (1996).
\bibitem{Vretenar2005PR} D. Vretenar,A.Afanasjev,G. Lalazissis, and P. Ring, Phys. Rep. 409, 101 (2005).
\bibitem{Meng2006PPNP} J. Meng, H. Toki, S. G. Zhou, S. Q. Zhang, W. H. Long, and L. S. Geng, Prog. Part. Nucl. Phys. 57, 470 (2006).
\bibitem{Meng2016} J. Meng, {\it Relativistic Density Functional for Nuclear Structure} (World Scientific, Singapore, 2016).

\bibitem{Niksic2009PRC}T. Nik\v{s}i\'{c}, Z. P. Li, D. Vretenar, L. Prochniak, J. Meng, and P. Ring, Phys. Rev. C 79, 034303 (2009).
\bibitem{Li2009PRCa}Z. P. Li, T. Nik\v{s}i\'{c}, D. Vretenar, J. Meng, G. A. Lalazissis, and P. Ring, Phys. Rev. C 79, 054301(2009).
\bibitem{Li2009PRCb}Z. P. Li, T. Nik\v{s}i\'{c}, D. Vretenar, and J. Meng, Phys. Rev. C 80, 061301(2009).
\bibitem{Li2010PRC}Z. P. Li, T. Nik\v{s}i\'{c}, D. Vretenar, P. Ring, and J.Meng, Phys. Rev. C 81, 064321(2010).
\bibitem{Li2011PRC}Z. P. Li, J.M. Yao, D. Vretenar, T. Nik\v{s}i\'{c}, H. Chen, and J. Meng, Phys. Rev. C 84, 054304(2011).
\bibitem{Li2012PLB}Z. P. Li, C. Y. Li, J. Xiang, J. M. Yao, and J. Meng, Phys. Lett. B 717, 470(2012).
\bibitem{Fu2013PRC}Y. Fu, H. Mei, J. Xiang, Z. P. Li, J. M. Yao, and J. Meng, Phys. Rev. C 87, 054305 (2013).
\bibitem{Lu2015PRC}K. Q. Lu, Z. X. Li, Z. P. Li, J. M. Yao, and J.Meng, Phys. Rev. C 91, 027304(2015).
\bibitem{Li2016JPG} Z. P. Li, T. Nik\v{s}i\'{c}, and D. Vretenar, J. Phys. G 43, 024005 (2016)
\bibitem{Niksic2011PPNP}T. Nik\v{s}i\'{c}, D. Vretenar, and P. Ring, Prog. Part. Nucl. Phys. 66,519(2011).
\bibitem{Prassa2012PRC}V. Prassa, T. Nik\v{s}i\'{c}, G. A. Lalazissis, and D. Vretenar, Phys. Rev. C 86, 024317 (2012).
\bibitem{Prassa2013PRC}V. Prassa, T. Nik\v{s}i\'{c}, and D. Vretenar, Phys. Rev. C 88, 044324(2013).
\bibitem{Xiang2016PRC}J. Xiang, J. M. Yao,  Y. Fu,  Z. H. Wang, Z. P. Li, and W. H. Long, Phys. Rev. C 93, 054324 (2016).
\bibitem{Prochniak2004NPA}L. Pr\'{o}chniak, P. Quentin, D. Samsoen, and J. Libert, Nucl. Phys. A 730, 59 (2004).
\bibitem{Libert1999PRC}J. Libert, M.Girod, and J.-P. Delaroche,Phys. Rev. C 60, 054301(1999).
\bibitem{Zhao2010PRC} P. W. Zhao, Z. P. Li, J. M. Yao, and J. Meng, Phys. Rev. C \textbf{82}, 054319 (2010).
\bibitem{Tian2009PLB} Y. Tian, Z. Y. Ma, and P. Ring, Phys. Lett. B \textbf{676}, 44 (2009).
\bibitem{Niksic2010PRC} T. Nik\v{s}i\'{c}, P. Ring, D. Vretenar, Y. Tian, and Z.~Y. Ma, Phys. Rev. C \textbf{81}, 054318 (2010).
\bibitem{Xiang2012NPA} J. Xiang, Z. P. Li, Z. X. Li, J. M. Yao, and J. Meng, Nucl. Phys. A \textbf{873}, 1 (2012).

\bibitem{Radich2015PRC}A. J. Radich, P. E. Garrett, J. M. Allmond et al., Phys. Rev. C 91, 044320(2015).
\bibitem{Rainovski2010PLB}G. Rainovskia, N. Pietralla, T.Ahn, L.Coquard, C. J. Lister, R. V. F. Janssens, M. P. Carpenter, S. Zhu, L. Bettermann, J. Jolie, W. Rother, R. V. Jolos, and V.Werner, Phys. Lett. B 683, 11 (2010).
\bibitem{NNDC} http://www.nndc.bnl.gov/chart/
\bibitem{Klemme1999PRC}T. Klemme, A. Fitzler, A. Dewald, S. Schell, S. Kasemann, R. Kuhn, O. Stuch, H. Tiesler, K. O. Zell, P. von Brentano, D. Bazzacco, F. Brandolini, S. Lunardi, C. M. Petrache, C.  RossiAlvarez, G. De Angelis, P. Petkov, and R. Wyss, Phys. Rev. C 60, 034301(1999).
\bibitem{Delaroche2010} J. -P. Delaroche, M. Girod, J. Libert, H. Goutte, S. Hilaire, S. P\'eru, N. Pillet, and G. F. Bertsch, Phys. Rev. C 81, 014303 (2010).
\bibitem{Kibedi2005ADNDT} T. Kib\'edi and R.H. Spear, At. Data Nucl. Data Tables 89, 77 (2005).
\end{thebibliography}
\end{document}